**Ion Chamber Collection Efficiencies for Proton Spot Scanning Calibration**

Keith M. Furutani*[1], Nicholas B. Remmes[2], Jon J. Kruse[2], Michael G. Herman[2]

Jiajian Shen[3], and Chris J. Beltran[1]

[1]*Department of Radiation Oncology, Mayo Clinic, Jacksonville, FL 32224*

[2]*Department of Radiation Oncology, Mayo Clinic, Rochester, MN 55905*

[3]*Department of Radiation Oncology, Mayo Clinic, Phoenix, AZ 85054*

*Corresponding author: furutani.keith@mayo.edu

**Purpose:** The motivation for this study was to determine the appropriate recombination model to correct for the ion chamber collection efficiency in the calibration of a Spot Scanning Proton Therapy System (Hitachi ProBeat V).  The accepted international protocol for calibration of proton therapy machines, TRS-398, specifies recombination correction for either pulsed or pulsed-scanned beams. The continuous approximation, not explicitly specified in the protocol, may in fact be the most appropriate correction.

**Methods:** Charge accumulation was measured under calibration conditions in the spread-out Bragg peak (SOBP) using the calibration bias as well as  a range of voltages from 10V to 500V and a Farmer-style ion chamber.   Collection efficiency was determined by extrapolating to infinite voltage.  Similar measurements were taken in an identical dose distribution with a much shorter spot duration.  The impact of each of the three models on calibration was then quantified using the TRS-398 protocol.

**Results:**  The collection efficiency for the standard calibration was determined to agree well with the prediction of a continuous beam recombination correction.  The standard calibration field was found to persistently agree with a continuous beam recombination correction for much lower operating biases.  The collection efficiency result for the short spot duration field did not agree with either the continuous or pulsed-beam correction.  Using the incorrect recombination model under the standard calibration conditions resulted in a 0.5% calibration difference.

**Conclusions:** We have determined that our spot scanning system would be most appropriately calibrated using a recombination correction with continuous beam model.  Physicists responsible for the calibration of such systems are advised to take measurements described here to correctly identify the applicable recombination model for their clinics.

INTRODUCTION

Models of recombination in ion chambers have been well described in the literature[1-6]. Moreover the explanation of the determination of the collection efficiency in gas ionization chambers for pulsed and continuous radiation specifically is lucidly described by Johns and Cunningham[7].

The accepted international protocol for calibration of proton therapy machines is TRS-398[8]. Continuous beam approximations for recombination as used in the AAPM TG-21[9] and AAPM TG-51 [10] protocols for photons are not indicated in Worksheet 10.8 of TRS-398. TRS-398 does require the user to make a determination for the recombination correction of either a pulsed or pulsed-scanned beam based on their knowledge of their proton source. The protocol excludes the continuous beam model from Worksheet 10.8 presumably because all proton therapy machines are pulsed. However, Lorin et al [11] have demonstrated that when the pulse duration is comparable to the ion transit time the recombination can be better modeled with a continuous beam approximation.

The purpose of this work is to report on the determination of the appropriate recombination model to use for our standard calibration field. The impact of using the pulsed, pulsed-scanned and continuous models is quantified using the TRS-398 protocol. Finally we investigated the effect of a shortened spot duration and/or increased recombination time as compared to the prediction of the different recombination models.

2. METHODS

2.A SPOT DURATION AND RECOMBINATION TIME

The recombination time in the ion chamber is voltage dependent, and for an air vented Farmer-style ion chamber operating at a voltage near saturation, recombination times are estimated to be approximately 0.1 ms using Equation 9-2 of Johns and Cunningham [7] below:

$$t_r = \frac{d^2}{V \times 3.6 \frac{cm\ cm}{sec Volt}} \tag{1}$$

The Hitachi ProBeat V Spot Scanning Proton Therapy System utilizes a synchrotron for proton acceleration. The bunch length of protons in the synchrotron and the time between bunches are both less than a microsecond. Therefore the synchrotron time structure is a factor of a thousand less than the ion chamber's recombination time. This time structure would therefore appear to be a continuous time structure within an extraction of an individual spot of protons. The Hitachi ProBeat V Spot Scanning Proton Therapy System uses a slow extraction method such that protons may be extracted at a rate of roughly $10^8$ protons per second, which is approximately 10 Monitor Units per second (MU/s). The minimum MU per beam spot is 0.003 MU, and the maximum is 5 MU, and with a steady beam current, the number of protons per spot is controlled by varying the spot duration. This spot duration can vary from about 0.1 ms to about 500 ms, which is comparable to or much longer than the recombination time for a typical Farmer chamber at a typical operating bias. Between beam spots within an energy layer is a latency time of approximately 2 ms so the time between spots is much longer than the recombination time. Clearly the spot duration is the time scale of importance for addressing recombination rate.

## 2.B MODELS OF RECOMBINATION

There are three different models to describe recombination rate $r$ based upon the time structure of the delivered beam compared to the recombination time $t_r$. The formula for the collection efficiency $f$ is simply the difference between unity and the recombination fraction: $f = 1 - rt_r$. The recombination correction factor $k_s = {}^1/_f$ and $k_s$ is referred to as $P_{ion}$ in TG-21 (9) and TG-51(10). In TG-21 $P_{ion}$ was plotted versus Q1/Q2 where Q1 is the charge collected at the reference bias voltage V1 and Q2 is the charge collected at half of the reference bias voltage V2=V1/2. These model predictions have been reproduced in Figure 2 in the Results section below. In the region where $P_{ion} \sim 1$ the predictions for Pulsed and Pulsed Scanning are observed to be very similar in Table II and IV of Reference (3). Therefore for the remainder of the discussion we will focus primarily on distinguishing between continuous and pulsed models. The difference between these models is based upon the assumed time structure of the delivered beam compared to the recombination time. This is illustrated in Figure 1 below.

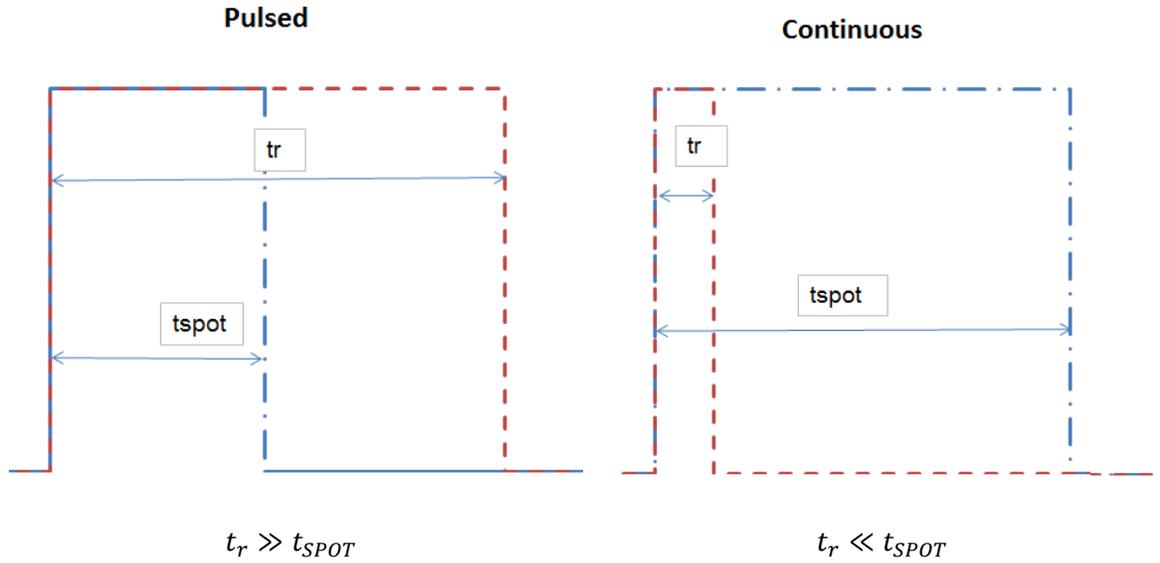

**Figure 1**: Difference between the Pulsed Recombination Model and Continuous Recombination Model assumptions showing the definition of recombination time $t_r$ and spot duration $t_{SPOT}$. Pulsed assumes $t_r \gg t_{SPOT}$ and Continuous assumes $t_r \ll t_{SPOT}$.

What is not immediately obvious is that a beam could be best described as $P_{ion,continuous}$ at a high operating bias where $t_r \ll t_{SPOT}$ but then best described as $P_{ion,pulsed}$ at a lower bias where now $t_r \gg t_{SPOT}$. To obtain the charge actually created $Q_{created}$ one extrapolates the charge collected $Q_{collected}$ at the highest, yet safe, operating bias to an infinite bias. Such an extrapolation could be confounded by a bias independent initial or columnar recombination of ions usually only found along high LET tracks. Estimates by Jaffe (12) for 2 MeV protons in air indicates that initial recombination would be negligible. We determined $Q_{created}$ by plotting $^1/_{Q_{collected}}$ versus $^1/_{V^2}$ for only the highest voltage readings and then extrapolating to infinite voltage. Note that if instead one plots of $^1/_{Q_{collected}}$ versus $^1/_V$ and then extrapolates to infinite voltage the same $^1/_{Q_{created}}$ is obtained to within 0.1%. The actual collection efficiency at the operating bias was then determined by $P_{ion,measured} = Q_{created}/Q_{collected}$. This result was then compared to that predicted by $P_{ion,continuous}$, $P_{ion,pulsed}$ and $P_{ion,pulsed\ scanning}$.

### 2.C IRRADIATION SETUP

The measurement chamber (a Farmer-like PTW 30013, Freiburg Germany) was operated in the standard configuration through a triaxial cable to a Fluke (Cleveland, OH) model 35040 electrometer with programmable bias selection through +/- 500V. The inner electrode was biased to collect to negative ions, the same conditions stated in the ion chamber calibration certificate.

The standard calibration field used at our clinic uniformly irradiates a 10cm x 10cm x 10cm SOBP extending from 10cm depth to 20cm depth with a dose of 2 Gy. The MU chambers are calibrated such that this 2 Gy irradiation to the one liter volume results from a delivery of 200 MU. The spots in the most distal layers of the pattern are delivered through multiple applications of 0.04 MU/spot, while the spots in the proximal layers require only a single application of approximately 0.005 MU. The chamber was placed at the center of the SOBP in an acrylic slab at a water-equivalent depth of 15 cm, and most of the dose in the center of the SOBP is delivered from spots with 0.02 MU/spot or more. Measurements were taken with a multichannel oscilloscope of the spot on and spot off logic signals during the delivery of the standard calibration field and the spot duration observed is generally between about 2 to 5 ms during the delivery of the most distal half of the treatment volume.

The TRS-398 calibration measurements were performed in the standard calibration field at 300 V and 400 V collection potentials. To study recombination dependence on spot duration a field was constructed (denoted the "minMU field") that irradiates the same volume to the same dose as the standard calibration field but using only spots between 0.003 MU/spot and 0.006 MU/spot. High dose distal layers were treated with multiple applications of low MU spots. Measurements of the spot duration on/off time for the minMU field show that the spot duration is generally around 0.5 ms. For these two fields measurements $Q_{collected}$ was obtained at a series of different voltages from 10 V to 500 V.

3. RESULTS AND DISCUSSION

Table 1 shows the data for the standard calibration field compared to the predictions from the Continuous Model, the Pulsed Model and the Scanned Pulsed Model. The standard calibration measurements in the proton beam are clearly consistent with the continuous radiation model. Calibration measurements at 300 and 400V in the standard calibration field show a 0.53% to 0.65% difference between the pulsed beam recombination model and $P_{ion,measured}$ whereas the difference is much less, 0.03% to 0.16%, between continuous beam recombination model and $P_{ion,measured}$. Measurements of the Hitachi ProBeat V Spot Scanning Proton Therapy System clearly demonstrate that a continuous beam recombination model is the most appropriate model for the standard calibration field. Significantly, the TRS-398 protocol does not explicitly suggest or allow for the continuous beam recombination model as part of proton calibration. Using the incorrect recombination model can result in calibration errors of 0.5% to 1% with commonly used measurement equipment and voltages

| Bias | Pion Measured | Continuous Model | Pulsed Model | Scanned Pulsed Model |
|------|---------------|------------------|--------------|----------------------|
| 300 | 1.0050 | 1.0034 | 1.0099 | 1.0107 |
| 400 | 1.0025 | 1.0028 | 1.0081 | 1.0091 |

**Table 1:** Recombination correction factor $P_{ion}$ for the standard field at the operating bias of 300 and 400 V determined using extrapolation to infinite voltage compared to the model predictions of Equation 14 and Equation 16 in TRS-398(8).

Figure 2 below shows the measurements with an operating bias from 10 to 500 Volts compared to the predictions of the three recombination models. The standard calibration measurements are clearly consistent with the continuous radiation model even for very low bias and inconsistent with the pulsed radiation model for recombination. This implies that even though $t_r$ is larger for the lower bias the standard calibration field measurements are still within the regime where $t_r \ll t_{SPOT}$. The minMU field has a much shorter spot duration and does appear to become more pulsed-like as the bias is decreased but the measurements do not lie clearly on either curve. Therefore it is possible that adjusting the delivery parameters so that spot duration is comparable to the recombination time $t_r \sim t_{SPOT}$ can result in a situation where neither continuous nor pulsed beam models are good fits. While not the case for the charge density, spot duration, operating bias and ion chamber used in our calibration it could be possible that another proton therapy system may inadvertently be calibrated where there is not a model prediction such as that demonstrated in these short spot duration and low bias measurements.

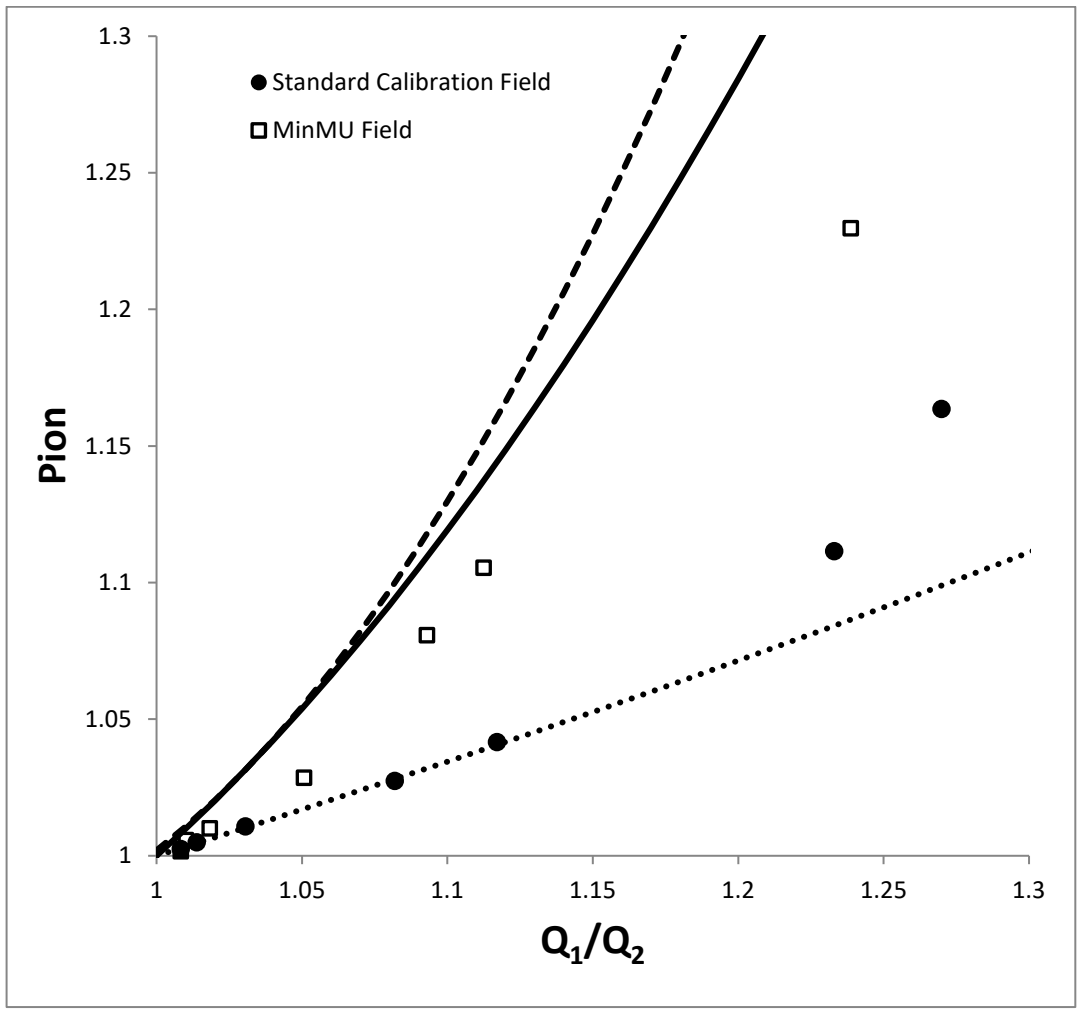

**Figure 2**: Plot of Recombination correction factor Pion versus Q1/Q2 where Q1 is the charge collected at the reference bias voltage V1 and Q2 is the charge collected at half of the reference bias voltage V2=V1/2. The dashed line is the prediction for a pulsed scanned beam. The solid line is the prediction for a pulsed beam. The dotted line is the prediction for a continuous beam. The circles are for the Standard calibration field. The squares are for the minMU field which has a much shorter spot duration than the standard calibration field.

## 4. CONCLUSIONS

Given that the measurements to determine which recombination model to use are straightforward to perform and given the rather large variation in proton therapy equipment and calibration fields, we recommend that a similar set of measurements as reported here should be part of commissioning of any new proton therapy facility. We recommend that these measurements should then be used to determine which is the best description of the recombination correction and if appropriate the continuous model recombination correction be used in spite of its omission from Worksheet 10.8 of TRS-398 protocol.

## ACKNOWLEDGEMENTS

The authors thank Mr. Nomura and Mr. Sakuma of Hitachi for their helpful cooperation